\documentclass[smallextended,referee,envcountsect]{svjour3} 
\usepackage{booktabs}
\usepackage{multirow}
\usepackage{diagbox}
\usepackage{subfigure} 
\usepackage{graphicx}
\usepackage{color}
\usepackage{algorithm}  
\usepackage{algorithmicx}  
\usepackage{algpseudocode}
\usepackage{array}
\usepackage{ulem}
\usepackage{float}
\usepackage{amsmath}
\usepackage{amssymb}
\usepackage{enumitem} 
\usepackage{hyperref}
\hypersetup{hidelinks,
	colorlinks=true,
	allcolors=black,
	pdfstartview=Fit,
	breaklinks=true}

\journalname{JOTA}

\newfloat{figtab}{htb}{fgtb}
\makeatletter
\newcommand\figcaption{\def\@captype{figure}\caption}
\newcommand\tabcaption{\def\@captype{table}\caption}

\newtheorem{myDef}{Definition} 
\newtheorem{myTheo}{Theorem}
\newtheorem{myprop}{Proposition} 
\newtheorem{myremar}{Remark}
\newtheorem{mylemma}{Lemma}
\newtheorem{assumption}{Assumption}

\begin{document}

\title{A Euclidean Distance Matrix Model for Convex Clustering}


\author{Z. W. Wang \and X. W. Liu \and Q. N. Li}

\institute{
	{\bf Communicated by Zaid Harchaoui} \\
	\rule[3ex]{3.82cm}{0.4pt}
	\vspace{-2em}
	\\
	Zhaowei Wang\at
	School of Xuteli,  Beijing Institute of Technology, 100081. Beijing. China. Academy of Mathematics and Systems Science, Chinese Academy of Sciences, 100190. Beijing. China. School of Mathematical Sciences, University of Chinese Academy of Sciences, 100049. Beijing. China \\
	\email{wangzhaowei23@mails.ucas.ac.cn}
	\and
	Xiaowen Liu \at
	School of Mathematics and  Statistics, Beijing Institute of Technology, 100081. Beijing. China \\
	\email{xiaowen.liu20@imperial.ac.uk}           
	\and
	Qingna Li \at
	Corresponding author. This author's research is supported by 12071032 and 12271526. \\
	School of Mathematics and Statistics/ Beijing Key Laboratory on MCAACI, Beijing Institute of Technology, 100081. Beijing. China \\
	\email{qnl@bit.edu.cn}        
}

\date{Received: 13 May 2021 / Accepted: 2 January 2025}

\maketitle

\begin{abstract}
	Clustering has been one of the most basic and essential problems in unsupervised learning due to various applications in many critical fields. 
	The recently proposed sum-of-norms (SON) model by Pelckmans et al. (2005), Lindsten et al. (2011) and Hocking et al. (2011) has received a lot of attention. 
	The advantage of the SON model is the theoretical guarantee in terms of perfect recovery, established by Sun et al. (2018). 
	It also provides great opportunities for designing efficient algorithms for solving the SON model. The semismooth Newton based augmented Lagrangian method by Sun et al. (2018) has demonstrated its superior performance over the alternating direction method of multipliers (ADMM) and the alternating minimization algorithm (AMA). 
	In this paper, we propose a Euclidean distance matrix model based on the SON model. Exact recovery property is achieved under proper assumptions. 
	An efficient majorization penalty algorithm is proposed to solve the resulting model. Extensive numerical experiments are conducted to demonstrate the efficiency of the proposed model and the majorization penalty algorithm.
\end{abstract}

\section{Introduction}
\label{intro}	

Clustering is one of the most basic and essential problems in unsupervised learning. It is to divide a group of data into several clusters so that the data in the same cluster are highly similar in some sense, whereas the data in different clusters are significantly different under some measurements. Clustering has been widely used in various applications in the fields of data analysis and machine learning.

Traditional clustering methods include the famous K-Means method, the hierarchical clustering, which may stick to a local minimum due to the nonconvexity of the models \cite{Sun}. They are also sensitive to the choices of initial points as well as the number of clusters $K$. Other methods like spectral clustering \cite{Fiedler1973}, which is a graph-based algorithm, can be quite unstable under different choices of the parameters for the neighborhood graphs \cite{von2007tutorial}.  In \cite{Hocking,Lindsten,Pelckmans}, a new clustering model called the sum-of-norm (SON) model was proposed, trying to tackle the above issues. It is a convex model and can be solved by alternating direction method of multipliers (ADMM) and alternating minimum algorithm (AMA) \cite{ChiandLange}. However, there is no theoretical guarantee in terms of exact recovery for SON for general weighted case, and AMA and ADMM are only restricted to the small scales of clustering. Recently, Sun et al. \cite{Sun} established the theoretical guarantee for the general weighted case, and a semismooth Newton based augmented Lagrangian method was proposed for the SON model, which can deal with large-scale problems. Very recently, Yuan et al. \cite{yuan2023randomly} considered a random dimension reduction approach for convex clustering, which greatly reduced the feature dimension of data.

On the other hand, the Euclidean Distance Matrix (EDM) based models for multidimensional scaling (MDS) have been proved to be successful tools to deal with problems arising from data visualization and dimension reduction \cite{BaiandQi,DingandQi,Qi2013,QiandYuan2014}. EDM models also find applications in sensor network localization \cite{BaiandQi,Qi2013A}, molecular conformation \cite{ZhaiandLi} and posture sensing for large manipulators \cite{Yao2020}. Compared with the traditional Semidefinite Programming (SDP)-based approaches for multidimensional scaling \cite{Biswas,Mordern,Cox,Dattorro,Toh2008}, the advantage of EDM based models deals with the EDM constraints via the characterization proposed by \cite{schoen} (see (\ref{eq:5}) forehead), which satisfies the constraint nondegeneracy. Moreover, fast numerical algorithms are proposed for different EDM models (such as semismooth Newton's method \cite{Qi2013,QiandYuan2014}, majorization penalty method \cite{QiXiuandZhou}, smoothing Newton's method \cite{LiandQi}). Very recently, Qi et al. \cite{QiXiuandZhou} proposed a new penalty technique to deal with the rank-constrained EDM model for penalized stress minimization with box constrains as well as robust EDM embedding \cite{ZhouandQi}. The new technique leads to the lower computational cost and therefore is able to deal with large-scale problems. Such technique is successfully applied to deal with the ordinal constrained EDM model \cite{LuandLi}.

Coming back to the clustering problem, it is actually highly related to distance, especially for the SON model. The success of EDM models in MDS motivates us to consider the following question: is it possible to build an EDM model for clustering? This is the main focus of our paper. The contributions of the paper are three folds. Firstly, we introduce an EDM-based model for clustering based on SON model, and establish the property of exact recovery. Secondly, inspired by the penalty technique in  \cite{QiXiuandZhou,ZhouandQi}, a fast majorization penalty algorithm is proposed to solve the resulting EDM model. Finally, we verify the efficiency of the EDM model and our algorithm by extensive numerical results.

The organization of the paper is as follows. In section \ref{sec:2}, we propose the EDM model for clustering, and the exact recovery property is established. In section \ref{sec:3}, we introduce the majorization penalty algorithm to solve the resulting model. In section \ref{sec:4}, we discuss how to solve the subproblem in an elementwise way. Numerical results are demonstrated in section \ref{numerical-experiments} to show the efficiency of the proposed model and the algorithm. Final conclusions are made in section \ref{sec:6}.

{\bf{Notations}}
We use $\| \cdot \| $ to denote the $\ell_2$ norm for vectors and Frobenius norm for matrices. We use $ diag(X) $ to denote the vector whose elements come from the diagonal elements of the matrix $X$. We use $Diag(x)$ to denote the diagonal matrix whose diagonal elements come from the vector $x$. Let $S^n$ denote the set of real symmetric matrices with size $n$ by $n$. We use $[n]$ to denote $\{1,\cdots,n\}$, and $|\Omega|$ to denote the number of elements in a set $\Omega$.

\section{EDM Model for Clustering}\label{sec:2}
In this part, we will propose the EDM model for clustering based on the SON model. In section \ref{sec:2.1}, we will briefly review the SON convex clustering model studied in \cite{Hocking}. In section \ref{sec:2.2}, we will reformulate the SON model by EDM, leading to the EDM model for clustering. In section \ref{sec2.3}, we establish the exact recovery property under proper conditions.
\subsection{The SON model}\label{sec:2.1}

Let $a_1, a_2,\cdots,a_n \in \mathbb{R}^{d}$ be the given data, where $n$ is the number of  observations and $d$ is the number of features.  The convex clustering model with general weights in terms of the sum of norms (SON) is given as follows \cite{Hocking} :
\begin{equation}\label{SON-p}
\min_{X \in \mathbb{R}^{d \times n}}\frac{1}{2} \sum^n_{i=1} \|x_i -a_i\|^2 + \gamma \sum_{i < j} \omega_{ij} \|x_i- x_j\|_{p}
\end{equation}
where $ \gamma >0 $ denotes the penalty parameter, $\| \cdot \|_p $ is the $\ell_p$ norm with $ p \geq 1$ and $ x_i $ is the "centroid" (the term used in \cite{Sun}, meaning the approximate one associated with $a_i$ but not the final cluster to which $a_i$ belongs) of the corresponding data $a_i$, and $ X := [x_1,x_2,...,x_n] \in \mathbb{R}^{d \times n} $. Here $\omega_{ij} = \omega_{ji} \geq 0$ are given weights on the input data $A :=[ a_1, a_2,\cdots,a_n] \in \mathbb{R}^{d \times n} $. If $\omega_{ij} = 1,\ i,j \in [n]$, (\ref{SON-p}) reduces to the typical convex clustering model studied in \cite{Hocking,Lindsten,Pelckmans}. In this paper, we consider the SON model (\ref{SON-p}) with $p=2$. That is,
\begin{equation}
\label{SON(A,d)}
	\min_{X \in \mathbb{R}^{d \times n}}\frac{1}{2} \sum^n_{i=1} \|x_i -a_i\|^2 + \gamma \sum_{i<j} \omega_{ij} \|x_i- x_j\|_{2} := h(X).
	\tag{SON($A,d$)}
\end{equation}
 
One can see that problem SON($A,d$) is a convex optimization problem. Let $ X^* := [x^*_1, x^*_2,...,x^*_n] \in \mathbb{R}^{d \times n} $ be the optimal solution of SON($A,d$). Then one can obtain the clustering in the following way : $a_i$ and $a_j$ belong to the same cluster if and only if $x^*_i =x^*_j$.

In \cite{Sun}, Sun et al. established the exact recovery result of (\ref{SON-p}) with general weights $\omega_{ij},\ i,j \in [n]$. Let $\{V_1,\cdots,V_K\}$ be a partition of $\{a_1,\cdots,a_n\}$. Define the index set $	I_\alpha:=\left\{i \mid a_i \in V_\alpha\right\}, \text { for } \alpha=[K]$, where $K$ is the number of clusters. Let $n_\alpha=\left|I_\alpha\right|$, and define 
\begin{equation*}
	\begin{aligned}
	& \mathbf{a}^{(\alpha)}=\frac{1}{n_\alpha} \sum_{i \in I_\alpha} a_i, \quad w^{(\alpha, \beta)}=\sum_{i \in I_\alpha} \sum_{j \in I_\beta} w_{i j}, \quad \forall~ \alpha, \beta \in [K], \\
	& w_i^{(\beta)}=\sum_{j \in I_\beta} w_{i j}, \quad \forall~ i \in [n], \beta \in [K],
	\end{aligned}
	\label{recovery-definition}
\end{equation*}
where $w_i^{(\beta)}$ can be interpreted as the coupling between point $a_i$ and the $\beta$-th cluster, 
and $w^{(\alpha, \beta)}$ as the coupling between the $\alpha$-th and $\beta$-th clusters. 
We call $\mathbf{a}^{(\alpha)}$ the clustering center 
because it is the average of the points in real cluster $\alpha$. 

The theoretical recovery guarantee of model (\ref{SON-p}) can be stated as follows.
\begin{myTheo}\label{gamma recovery}
	{\rm{\cite[Theorem 5]{Sun}}} Given input data $A=\left[a_1, a_2, \cdots, a_n\right] \in \mathbb{R}^{d \times n}$ and its partitioning $\mathcal{V}=$ $\left\{V_1, V_2, \ldots, V_K\right\}$. Assume that all the clustering centers $\left\{\mathbf{a}^{(1)}\right.$, $\mathbf{a}^{(2)}, \ldots$, $\left.\mathbf{a}^{(K)}\right\}$ are distinct. Denote the optimal solution of (\ref{SON-p}) by $\left\{x_i^*\right\}$ and define the map $\psi\left(a_i\right)=x_i^*$ for $i=1, \ldots, n$. The following results hold.
	
	\begin{itemize}
		\item[(i)] Let
		$
		\mu_{i j}^{(\alpha)}:=\sum_{\beta=1, \beta \neq \alpha}^K\left|w_i^{(\beta)}-w_j^{(\beta)}\right|, \quad i, j \in I_\alpha, \alpha \in [K]
		$. Assume that $w_{i j}>0$ and $n_\alpha w_{i j}>\mu_{i j}^{(\alpha)}$ for all $i, j \in I_\alpha, \alpha \in [K]$. Let
		$$
		\begin{aligned}
		& \gamma_{\min }:=\max _{1 \leq \alpha \leq K} \max _{i, j \in I_\alpha}\left\{\frac{\left\|\mathbf{a}_i-\mathbf{a}_j\right\|_2}{n_\alpha w_{i j}-\mu_{i j}^{(\alpha)}}\right\}, \\
		& \gamma_{\max }:=\min _{1 \leq \alpha<\beta \leq K}\left\{\frac{\left\|\mathbf{a}^{(\alpha)}-\mathbf{a}^{(\beta)}\right\|_2}{\frac{1}{n_\alpha} \sum_{1 \leq l \leq K, l \neq \alpha} w^{(\alpha, l)}+\frac{1}{n_\beta} \sum_{1 \leq l \leq K, l \neq \beta} w^{(\beta, l)}}\right\} .
		\end{aligned}
		$$
		If $\gamma_{\min }<\gamma_{\max }$ and $\gamma$ is chosen such that $\gamma \in\left[\gamma_{\min }, \gamma_{\max }\right)$, then the map $\psi$ perfectly recovers $\mathcal{V}$.
		
		\item[(ii)]
		If $\gamma$ is chosen such that
		\begin{equation*}
		\gamma_{\min } \leq \gamma<\max _{1 \leq \alpha \leq K} \frac{n_\alpha\left\|\mathbf{c}-\mathbf{a}^{(\alpha)}\right\|_2}{\sum_{1 \leq \beta \leq K, \beta \neq \alpha} w^{(\alpha, \beta)}}
		\end{equation*}
		
		where $\mathbf{c}=\frac{1}{n} \sum_{i=1}^n \mathbf{a}_i$, then the map $\psi$ perfectly recovers a non-trivial coarsening of $\mathcal{V}$.
	\end{itemize}
\end{myTheo}

\begin{myremar}
	Theorem \ref{gamma recovery} (i) implies that if $\gamma \in \left[\gamma_{\min },\gamma_{\max }\right)$, the solution returned by SON($A,d$) recovers the underlying partition of input data $A$.
\end{myremar}

\subsection{EDM model}\label{sec:2.2}

In this part, we will reformulate problem SON($A,d$) as an EDM model. To that end, we start with the definition of EDM as well as the related properties.
\begin{myDef}\label{def1}
	A matrix $D \in S^q$ is an EDM if there exists a set of points $p_1,...,p_q \in \mathbb{R}^r \; (r \leq n-1)$ such that $D_{ij} =\|p_i - p_j\|^2,\ i, j \in [q]$. Here $r$ is referred to as the embedding dimension. 
\end{myDef}

The following characterization of EDM is given by \cite{schoen}.
\begin{myprop}
	A matrix $D$ is an EDM if and only if the following holds:
	\begin{equation}\label{eq:5}
	diag(D) = 0, \quad -D \in \mathcal{K}^q_+,
	\end{equation}
	where $\mathcal{K}^q_+$ is the conditional positive semidefinite cone given by
	\begin{equation*}\label{eq:6}
	\mathcal{K}^q_+ := \{ X \in S^q \;|\; {x}^\top X {x} \ge 0,\  \forall \ x \in \mathbb{R}^q \  \text{satisfying}\  {x}^\top {e} = 0  \}.
	\end{equation*}
	
	Here $e \in \mathbb{R}^q$ is the vector with all elements one.
\end{myprop}

Moreover, define $J \in S^q$ as $ J= I-\frac{1}{q}{ee}^\top $, where $ I \in S^q$ is the identity matrix. The embedding dimension $r$ is calculated by $ r=\text{rank}(JDJ) $.

Let $ D \in S^q $ be an EDM. One can use the following Classical Multidimensional Scaling (CMDS) to obtain a set of points $ y_1,\cdots,y_q \in \mathbb{R}^r $ that generates $D$. That is,
\begin{equation}
	-\frac{1}{2}JDJ = P\Lambda P^\top,\ [y_1,\cdots,y_q] = Diag(\lambda_1^{\frac{1}{2}},\cdots, \lambda_r^{\frac{1}{2}}) \cdot \hat{P}^\top.
	\label{CMDS-1}
\end{equation}
Here $\Lambda = Diag(\lambda_1,\cdots,\lambda_q),\ $ $\lambda_1,\cdots,\lambda_q$ are the eigenvalues arranged in nonincreasing order, with $r=\text{rank}(JDJ)$, and $P$ is the matrix with corresponding eigenvectors as columns. $\hat{P} \in \mathbb{R}^{q \times  r}$ contains the eigenvectors corresponding to the positive eigenvalues $\lambda_1, \cdots, \lambda_r$.

Equipped with the above preliminaries about EDM, we will interpret model SON($A,d$) from the EDM point of view. We try to use an EDM $D$ as the variable of the model. Consequently, we take the given points $ a_1,\cdots,a_n \in \mathbb{R}^d $ and the unknown points $ x_1,\cdots,x_n \in \mathbb{R}^d$ as those points $ Z = $ $[a_1,$ $\cdots,a_n,$ $x_1,\cdots,x_n]$ $:=$ $[z_1,\cdots,z_{2n}]$ $\in \mathbb{R}^{d \times 2n} $. Define $D \in S^{2n}$ to be the EDM generated by the points $z_1,\cdots,z_{2n}$. That is,
\begin{equation*}\label{eq:9}
D_{ij} =  \| z_i - z_j \|^{2}, \quad i,j \in [2n].
\end{equation*}
Specifically, $D_{ij}$ is defined by
\begin{equation}\label{eq:10}
D_{ij} =  \| a_i - a_j \|^{2}, \  i,j \in [n];
\end{equation}
\begin{equation}\label{eq:11}
D_{i \; n+i} =  \| a_i - x_i \|^{2}, \  i \in [n];
\end{equation}
\begin{equation}\label{eq:12}
D_{n+i \; n+j} =  \| x_i - x_j \|^{2}, \  i,j \in [n].
\end{equation}

Combining the objective function in model SON($A,d$) and (\ref{eq:11}), (\ref{eq:12}), we reach the following equivalent objective function in terms of EDM matrix $D$
\begin{equation*}\label{eq:13}
f(D):=\frac{1}{2} \sum_{i=1}^{n} D_{i~ n+i} + \gamma \sum_{i<j} \omega_{ij} \sqrt{D_{n+i~ n+j}}.
\end{equation*}

Coming to the constraints, we require that $D \in S^{2n}$ is an EDM with the embedding dimension $r$. Moreover, since $a_1,\cdots,a_n$ are available , $D_{ij}$ is known for $i,j \in [n]$. Consequently, we reach the following EDM model
\begin{equation}\label{model with D}
	\begin{aligned}
	\min_{D \in S^{2n}} & f(D)  \nonumber \\
	\hbox{s.t.}\; &diag(D) = 0, \nonumber \\
	&D_{ij} = \| a_i - a_j \|^{2}, \quad i,j \in [n], \nonumber \\
	&-D \in \mathcal{K}_{+}^{2n}(r),
	\end{aligned}
	\tag{EDM($r$)}
\end{equation}
where $r$ is given, and $\mathcal{K}_{+}^{2n}(r)$ is the conditional positive semidefinite cone with rank-$r$ cut defined by $ \mathcal{K}_{+}^{2n}(r):=\{ X \in \mathcal{K}^{2n}_+  \ | \  \text{rank}(JXJ) \le r \}$.

Let $H \in S^{2n}$ and $ W \in S^{2n} $ be defined by
\begin{equation*}\label{eq15}
\begin{aligned}
&H_{ij} = \left\{ 
	\begin{array}{lr}
	\frac{1}{4}, & \text{if} \ j=i+n,\ i \in [n]; \; \text{or}\  i=n+j,\ j \in [n], \\
	0,           & \hbox{otherwise}.
	\end{array} \right. \\	
&W_{ij} = \left\{
	\begin{array}{lr}
	\frac{1}{2}\omega_{i-n,j-n},\ \ & n<i<j\le 2n,\\
	\frac{1}{2}\omega_{j-n,i-n},\ \ & n<j<i\le 2n,\\
	0,     & \hbox{otherwise}. 
	\end{array} \right.
\end{aligned} 
\end{equation*}

Let $\sqrt{D}$ be defined by $(\sqrt{D})_{ij} = \sqrt{D_{ij}},\ i,j\in [2n]$. Moreover, define
\begin{equation}\label{eq:0}
B:= \left\{D \in S^{2n} \;|\; diag(D)=0,\ D_{ij} = \| a_i-a_j \|^2,\ i,j\in [n] \right\}.
\end{equation}
EDM($r$) can be equivalently reformulated as the following compact form:
\begin{equation}
\begin{aligned}
	\min_{D \in S^{2n}}& f(D) = \langle H,D \rangle + \gamma \langle W,\sqrt{D} \rangle  \\
	\hbox{s.t.}&\ -D \in \mathcal{K}_{+}^{2n}(r),\ D \in B. 
\end{aligned}
\label{compactEDM}
\end{equation}

\subsection{Exact recovery property of EDM($r$)}\label{sec2.3}

In this part, we discuss the exact recovery property of EDM($r$), which is heavily based on the relationship between the optimal solution of SON($A,d$) and EDM($r$). To that end, we first establish the so-called one-to-one correspondence between the feasible points of SON($A,d$) and the feasible matrix of EDM($r$). 

Consider the case that $r=d$. For any feasible point $x_1,\cdots,x_n \in \mathbb{R}^d$ of SON($A,d$), the EDM matrix $D$ generated by $a_1,\cdots,a_n, x_1,\cdots,x_n$ is obviously a feasible matrix of \ref{model with D}. Now we will show how to obtain a feasible solution of SON($A,d$) by a feasible matrix $D$ of EDM($d$). Let $D$ be a feasible matrix of \ref{model with D}. Conducting CMDS (\ref{CMDS-1}) on such $D$, one can obtain $y_1,\cdots,y_{2n} \in \mathbb{R}^d$ which generates $D$. Recall that $D$ satisfies (\ref{eq:10}), which implies that $y_1,\cdots,y_{n} \in \mathbb{R}^d$ generates the same EDM as $a_1,\cdots,a_n \in \mathbb{R}^d$. Therefore, there exists a linear mapping $L : \mathbb{R}^d \rightarrow \mathbb{R}^d$ and $b \in \mathbb{R}^d$ such that the following holds\footnote{One way to find such $L$ and $b$ is the well-known Procrustes process \cite{gower2004}.}
\begin{equation*}
	\left[y_1,\cdots,y_{n}\right] = L \left[a_1,\cdots,a_n\right] + b.
\end{equation*}
Let $x_1,\cdots,x_n \in \mathbb{R}^d$ be defined by
\begin{equation*}
\label{linear-map}
\left[x_1,\cdots,x_{n}\right] = L \left[y_{n+1},\cdots,y_{2n}\right] + b,
\end{equation*}
which gives a set of feasible points of SON($A,d$) corresponding to the feasible matrix $D$ of \ref{model with D}.

To discuss a more general case that $r \le d$, we make the following assumption.
\begin{assumption}
	Let $s \le \min \left(d,n\right)$ be the rank of $A=\left[a_1,\cdots,a_n\right] \in \mathbb{R}^{d \times n}$. Let $V \in S^n$ be the EDM generated by $a_1,\cdots,a_n \in \mathbb{R}^d$, and $\hat{a}_1,\cdots,\hat{a}_n \in \mathbb{R}^s$ is generated by applying CMDS (\ref{CMDS-1}) to V.
	\label{assumption-main}
\end{assumption}

For a feasible matrix $D$ of EDM($s$), we can conduct CMDS (\ref{CMDS-1}) to obtain $y_1,\cdots,y_{2n} \in \mathbb{R}^s$. Notice that both $\hat{a}_1,\cdots,\hat{a}_n$ and $y_1,\cdots,y_{n}$ generate the same EDM $V$. There exists a linear mapping $\hat{L} \in \mathbb{R}^s \rightarrow \mathbb{R}^s$ and $\hat{b} \in \mathbb{R}^s$ such that the following holds $$
\left[y_1,\cdots,y_{n}\right] = \hat{L} \left[\hat{a}_1,\cdots,\hat{a}_n\right] + \hat{b}.
$$ Similarly, one can obtain $\hat{x}_1,\cdots,\hat{x}_n \in \mathbb{R}^s$ by $$
\left[\hat{x}_1,\cdots,\hat{x}_n \right] = \hat{L} \left[\hat{a}_1,\cdots,\hat{a}_n\right] + \hat{b}.
$$ Therefore, for a feasible point $D$ of EDM($s$), one can obtain points $\hat{x}_1,\cdots,\hat{x}_n$ which are feasible to SON($\hat{A},s$) with $\hat{A} := \left[\hat{a}_1,\cdots,\hat{a}_n\right] \in \mathbb{R}^{s \times n}$.

To summarize, we have the following lemma.
\begin{mylemma}
	\label{lemma-summarize}
	\begin{itemize}
		\item[(i)] If $r =d$, there is a one-to-one correspondence between feasible points of SON($A,d$) and \ref{model with D}.
		\item[(ii)] If Assumption 1 holds, there is a one-to-one correspondence between feasible points of SON($\hat{A},s$) and EDM($s$).
	\end{itemize}
\end{mylemma}

\begin{myremar}
	In fact, Lemma \ref{lemma-summarize} (i) is a special case of Lemma \ref{lemma-summarize} (ii), where $s=d$.
\end{myremar}

Moreover, the following result holds.
\begin{myTheo}\label{theorem-1}
	(i) If $r = d$, SON($A,d$) is equivalent to EDM($d$).
	
	(ii) If Assumption 1 holds, SON($\hat{A},s$) is equivalent to EDM($s$).
\end{myTheo}
\begin{proof}
	(i) Note that for each feasible point $X \in \mathbb{R}^{d \times n}$ of SON($A,d$) and its corresponding feasible point $D \in S^{2n}$ of \ref{model with D},  it holds that $f(D) = h(X)$. Therefore, problem \ref{model with D} is equivalent to SON($A,d$).
	
	(ii) If Assumption \ref{assumption-main} holds, by the similar argument as above, one can also obtain that SON($\hat{A},s$) is equivalent to EDM($s$). \qed
\end{proof}

In terms of global minimizers of the two problems SON($A,d$) and EDM($r$), we have the following result. 

\begin{myTheo}
	(i) If $r = d$, there is a one-to-one correspondence between global minimizers of SON($A,d$) and EDM($d$). Moreover, the global minimizer of EDM($d$) is unique.
	(ii) If Assumption 1 holds, there is a one-to-one correspondence between the global minimizers of SON($\hat{A},s$) and EDM($s$). Moreover, the global minimizer of EDM($s$) is unique.
	\label{minimizer-unique}
\end{myTheo}
\begin{proof}
	(i) First we will show that there is a one-to-one correspondence between the global minimizers of SON($A,d$) and that of EDM($d$). Let $D^*$ be a global minimizer of EDM($d$). By Lemma \ref{lemma-summarize}, there exists $X^* = \left[x_1^*,\cdots,x_n^*\right] \in \mathbb{R}^{d \times n}$ such that $x_1^*,\cdots,x_n^*$ generate $D^*$. For any $X = \left[x_1,\cdots,x_n\right] \in \mathbb{R}^{d \times n}$, the EDM $D$ generated by $a_1,\cdots,a_n,x_1,\cdots,x_n$ is a feasible point of EDM($d$) and it also satisfies $f(D) \ge f(D^*)$. It holds that $h(X) = f(D) \ge f(D^*) = h(X^*)$, which implies that $X^*$ is the global optimal solution of SON($A,d$). 
	
	Conversely, let $X^*$ be the global optimal solution of SON($A,d$). Let $D^*$ be the EDM generated by $a_1,\cdots,a_n,x_1^*,\cdots,x_n^*$. For any feasible matrix $D \in S^{2n}$ of EDM($d$), by Lemma 1 (i), one can obtain $x_1,\cdots,x_n$ together with $a_1,\cdots,a_n$ that generate $D$. Such $D$ is a feasible matrix of EDM($d$). Moreover, we have $f(D) = h(X) \ge h(X^*) = f(D^*)$. Therefore, $D^*$ is a global minimizer of EDM($d$). 
	
	Overall, there is a one-to-one correspondence between the global minimizer of SON($A,d$) and EDM($d$). Finally, note that SON($A,d$) is strongly convex, the global minimizer of SON($A,d$) is unique. Therefore, the global minimizer of EDM($d$) is unique as well. Similar argument can be applied to show (ii).	\qed
\end{proof}
	
To summarize, for the exact recovery property of \ref{model with D} in terms of parameter $\gamma$ and $r$, we have the following result.
\begin{myTheo}
	\begin{itemize}
		\item[(i)] If $r = d$, by choosing $\gamma$ as in Theorem \ref{gamma recovery} (i), the global minimizer of EDM($d$) recovers the partition $\mathcal{V}$.
		\item[(ii)] If Assumption \ref{assumption-main} holds, let $\hat{\gamma}_{\min }$ and $\hat{\gamma}_{\max }$ be calculated in the same way as $\gamma_{\min }$ and $\gamma_{\max }$ in Theorem \ref{gamma recovery} (i), but with $a_1,\cdots,a_n \in \mathbb{R}^d$ replaced by $\hat{a}_1,\cdots,\hat{a}_n \in \mathbb{R}^s$. Then the global minimizer of EDM($s$) recovers the partition $\mathcal{V}$.
	\end{itemize}
	\label{EDM-recover}
\end{myTheo}
\begin{proof}
	(i) By Theorem \ref{minimizer-unique}, the global minimizer of EDM($d$) denoted by $D^*$ is unique, and the global optimal solution of SON($A,d$) denoted by $x^*_1,\cdots,x^*_n \in \mathbb{R}^d$ can be derived from $D^*$. Together with Theorem \ref{gamma recovery} (i), one can obtain the mapping $\psi$, which perfectly recovers $\mathcal{V}$.
	
	(ii) If Assumption \ref{assumption-main} holds, the result can be shown in the same way by replacing $a_1,\cdots,a_n \in \mathbb{R}^d$ by $\hat{a}_1,\cdots,\hat{a}_n \in \mathbb{R}^s$. That is, the global minimizer of EDM($s$) recovers the partition of $\{\hat{a}_1,\cdots,\hat{a}_n\}$. By Assumption 1, the partition of $\{\hat{a}_1,\cdots,\hat{a}_n\}$ is the same as that of $\{a_1,\cdots,a_n\}$. Therefore, the global minimizer of EDM($s$) recovers the partition $\mathcal{V}$ of $\{a_1,\cdots,a_n\}$.	\qed
\end{proof}

\begin{myremar}
	By Theorem \ref{EDM-recover}, we can also conduct similar argument to the case $s < r < d$. That is, for $s < r < d$, if $\gamma$ is chosen properly as in Theorem \ref{gamma recovery} (i), the global minimizer of EDM($r$) recovers the partition $\mathcal{V}$ of $\{a_1,\cdots,a_n\} \in \mathbb{R}^{d}$. 
\end{myremar}


\section{Majorization Penalty Method for EDM(r)}\label{sec:3}
In this part, we will propose the majorization penalty method to solve the EDM model (\ref{compactEDM}). 
The majorization penalty method was initially proposed to deal with the rank constrained nearest correlation matrix problem \cite{YandGao}. Then it was used to solve other nonconvex problems such as rank-constrained nearest EDM problem \cite{LuandLi,QiandYuan2014,QiXiuandZhou,ZhouandQi} and nearest correlation matrix problem with factor structure \cite{2011Block}. Very recently, it was applied to solve the sparse constrained support vector machine model \cite{LuandLi}. Below, we adopt the framework of majorization penalty method discussed in \cite{ZhouandQi}.

Recall the EDM model in \ref{model with D}. To deal with the nonconvex constraint $ -D \in \mathcal{K}_{+}^{2n}(r) $, we define the function $ g:S^{2n}  \xrightarrow{} \mathbb{R} $ by 
\begin{equation*}
g(A):=\frac{1}{2}\text{dist}^2(-A,\mathcal{K}_+^{2n}(r)),\ \forall \ A \in S^{2n}
\end{equation*}
where
\begin{equation*}\label{eq:17}
\text{dist}(A,\mathcal{K}_+^{2n}(r)):= \min\{ \| A-X \| \;|\; X \in \mathcal{K}_+^{2n}(r) \}.
\end{equation*}

Due to the definition above, $ -D \in \mathcal{K}_+^{2n}(r) $ if and only if $ g(D) = 0 $. Problem \ref{model with D} is equivalent to the following problem:
\begin{equation}
	\min_{D \in S^{2n}} f(D),\ \hbox{s.t.} \; D \in B,\ g(D) = 0.  \\
\label{model with g}
\end{equation}

To deal with the nonconvex constraint $g(D) = 0$, we penalize it, and solve the resulting penalized problem as follows:
\begin{equation}\label{model with rhog}
	\min_{D \in S^{2n}} f(D)+\rho g(D), \ \hbox{s.t.} \ D \in B,
\end{equation}
where $\rho > 0$ is a penalty parameter. To solve (\ref{model with rhog}), $g(D)$ is still not easy to tackle. Therefore, we follow the idea of majorization method. 
A function $g_m(\cdot, A)$ is said to be a majorization of $g(\cdot)$ at $A \in S^{2n}$ if it satisfies the following conditions
\begin{equation}
g_m(A,A) = g(A), \ g_m(D,A) \ge g(D),\ \forall \ D \in S^{2n}.
\end{equation}

Due to the above definition of majorization, at each iteration $D^\zeta$, we can construct a majorization function $g_m(D,D^\zeta)$, and solve the following problem
\begin{equation}\label{eq:20}
	\min_{D \in S^{2n}} f(D)+\rho g_m(D,D^\zeta),\ \hbox{s.t.} \  D \in B.
\end{equation}

The remaining question is to derive a majorization function $g_m(D,D^\zeta)$. Here we use the same majorization function as derived in \cite{QiXiuandZhou}, which is given below. See \cite{QiXiuandZhou} for more details.


Let $\Pi_{\mathcal{K}_+^{2n}(r)}^{B}(A)$ be the solution set of problem (\ref{eq:17}). That is,
\begin{equation*}
{\Pi}_{\mathcal{K}_+^{2n}(r)} ^B (A) = \mathop{\arg\min_{D \in S^{2n}}} \{ \| A-D \|, \ D \in \mathcal{K}_+^{2n}(r) \}.
\end{equation*}	
Let $ \Pi_{\mathcal{K}_+^{2n}(r)}(A) \in   \Pi_{\mathcal{K}_+^{2n}(r)}^B (A) $. That is, $ \Pi_{\mathcal{K}_+^{2n}(r)}(A)$ is one element in set $ \Pi_{\mathcal{K}_+^{2n}(r)}^B (A)$. Then at point $A \in S^{2n}$, there is
\begin{equation}\label{eq:21}
g(D) \le \frac{1}{2} \| D \|^2 - \frac{1}{2}\| {\Pi}_{\mathcal{K}_+^{2n}(r)}  (-A) \|^2 + \langle {\Pi}_{\mathcal{K}_+^{2n}(r)}(-A), D-A  \rangle := g_m(D,A).
\end{equation}
The function $g_m(\cdot,A)$ can be viewed as a majorization function of $g(\cdot)$ at $A \in S^{2n}$.

Now we are ready to present the majorization penalty method for (\ref{model with g}).

\begin{algorithm}[H]
	\caption{Majorization Penalty Method for (\ref{model with g})}
	\label{alg1}
	\begin{algorithmic}[S]
		\State S0: Given $\rho > 0,\ H\ and\ W \in S^{2n},\  r > 0 $, initial point $D^0;\ \zeta:=0$.
		\State S1: Solve the subproblem (\ref{eq:20}) with $g_m(D,D^\zeta)$ defined as in (\ref{eq:21}) to obtain $D^{\zeta+1}.$
		\State S2: If the stopping criterion is satisfied, stop; otherwise, $\zeta = \zeta+1$, go to S1.
	\end{algorithmic}
\end{algorithm}

We have the following remarks regarding Algorithm \ref{alg1}.
\begin{myremar}
	$\Pi_{\mathcal{K}_+^{2n}(r)}(A)$ can be calculated in the following way. Let $A \in S^{2n}$, admit the spectral decomposition as $ A=\sum_{i=1}^{2n} \lambda_i {p}_i {p}_i^\top $, where $\lambda_1 \ge \cdots \ge \lambda_{2n}$ are the eigenvalues of $A$ and ${p}_1,\cdots, {p}_{2n}$ are the corresponding eigenvectors. Define the principle-component-analysis (PCA) -style matrix truncated at $r$ as
	\begin{equation}\label{new1}
	\hbox{PCA}^+_r(A) = \sum_{i=1}^{r} \max\{0,\lambda_i\}{p}_i {p}_i^\top.
	\end{equation}
	Then $\Pi_{\mathcal{K}_+^{2n}(r)}(A)$ can be calculated as 
	\begin{equation}\label{new2}
	\Pi_{\mathcal{K}_+^{2n}(r)}(A) = \hbox{PCA}_r^+(JAJ) + (A-JAJ).
	\end{equation}  
\end{myremar} 

Note that in (\ref{new1}) only the first $r$ largest eigenvalues are involved. Therefore, when we calculate $\Pi_{\mathcal{K}_+^{2n}(r)}(A)$ by (\ref{new2}), we only need partial spectral decomposition rather than the full spectral decomposition. It will reduce the computational cost. Moreover, if $r$ is much smaller than $2n$, such advantage will bring more significant reduction in computational cost.

In terms of the convergence of our method, we have the following result, which comes from Theorem 3.2 in \cite{QiXiuandZhou}. 
Let $D$ be a feasible point of EDM($r$). Let $\rho_\epsilon = \frac{f(D)}{\epsilon}$, where $\epsilon > 0$.  
\begin{myTheo}
	{\rm{\cite[Theorem 3.2]{QiXiuandZhou}}} Let $\epsilon>0$ be given. For any $\rho \geq \rho_\epsilon$, let $D_\rho^*$ be an optimal solution of (\ref{model with rhog}) and $D^*$ is an optimal solution of EDM($r$). Then $D_\rho^*$ must be $\epsilon$-optimal. That is,
	$$ D_\rho^* \in B, \quad g\left(D_\rho^*\right) \leq \epsilon \quad \text { and } \quad f\left(D_\rho^*\right) \leq f\left(D^*\right). $$
	\label{Theo5Appendix}
\end{myTheo}

Proof can be seen in the Appendix.


\section{Algorithm for Solving Subproblem (\ref{eq:20})}\label{sec:4}
In this section we discuss how to solve the subproblem (\ref{eq:20}). Firstly, we will simplify it as shown in section \ref{sec:4.1}. Then we will discuss the solutions of simplified subproblem due to different cases, as shown in section \ref{sec:4.2}.

\subsection{Simplifying subproblem (\ref{eq:20})}\label{sec:4.1}
With $g_m(D,D^\zeta)$ defined as in (\ref{eq:21}), the objective function in (\ref{eq:20}) can be simplified as follows.

\begin{equation*}
\begin{aligned}
f(D) + \rho g_m(D,D^\zeta) = &\langle H,D \rangle + \gamma \langle W,\sqrt{D} \rangle + \frac{\rho}{2} \left\| D \right\|^2 \\ 
&+ \rho \left\langle {\Pi}_{\mathcal{K}_+^{2n}(r)} (-D^\zeta) ,  D-D^\zeta \right\rangle - \frac{\rho}{2} \left\| {\Pi}_{\mathcal{K}_+^{2n}(r)} (-D^\zeta) \right\| \\
:= & \frac{\rho}{2} \|D\|^2 + \left\langle D,H+\rho {\Pi}_{\mathcal{K}_+^{2n}(r)} (-D^\zeta) \right\rangle + \gamma \left\langle W,\sqrt{D} \right\rangle + C \\ 
:= & \frac{\rho}{2} \left\|D \right\|^2 + \left\langle D,\hat{D}^\zeta \right\rangle + \gamma \left\langle W,\sqrt{D} \right\rangle + C \\
= & \sum_{i=1}^{2n} \sum_{j=1}^{2n}
\left( \frac{\rho}{2}D_{ij}^2 + D_{ij} \hat{D}_{ij}^\zeta + \gamma W_{ij} \sqrt{D_{ij}} \right) +C,
\end{aligned}
\end{equation*}
where $C=-\frac{\rho}{2} \| \Pi_{\mathcal{K}_+^{2n}(r)}(-D^\zeta) \|^2 - \rho \langle \Pi_{\mathcal{K}_+^{2n}(r)}(-D^\zeta), D^\zeta \rangle $ is a constant with respect to $D$, and $\hat{D}^\zeta:= H + \rho \Pi_{\mathcal{K}_+^{2n}(r)}(-D^\zeta)$. 

Problem (\ref{eq:20}) then reduces to the following form
\begin{equation}\label{eq:22}
	\min_{D \in S^{2n}} \  \sum_{i=1}^{2n} \sum_{j=1}^{2n}
	\left( \frac{\rho}{2}D_{ij}^2 + D_{ij} \hat{D}_{ij}^\zeta + \gamma W_{ij} \sqrt{D_{ij}} \right),  \;  \hbox{s.t.} \ D \in B.
\end{equation}

Recall the set $B$ defined as in (\ref{eq:0}). Consequently, for $i,j \in [n]$, $D_{ij}$ is fixed as $\|a_i-a_j \|^2$. Moreover, $D_{ii} = 0,\ i \in [2n]$. Problem (\ref{eq:22}) then reduces to the following one-dimensional optimization problem
\begin{equation}\label{eq:23}
	\min \ \frac{1}{2} \left(D_{ij} + \frac{\hat{D}_{ij}^{\zeta}}{\rho} \right)^2 + \frac{\gamma}{\rho} W_{ij} \sqrt{D_{ij}},  \;  \hbox{s.t.} \  D_{ij} \ge 0,
\end{equation}
where $i<j$ and $i,j$ are not in $[n]$ simultaneously. Note that the nonnegative constraint $D_{ij} \ge 0$ comes from the necessary condition that $D \in S^{2n}$ must be an EDM.	Next, we consider solving the general form of (\ref{eq:23}) (here $a=-\frac{\hat{D}_{ij}^k}{\rho}$, $b=\frac{\gamma}{\rho}W_{ij} \ge 0$)
\begin{equation}\label{eq:24}
\min_{\alpha \ge 0} \ \frac{1}{2}(\alpha-a)^2 + b\sqrt{\alpha}:=\varphi(\alpha).
\end{equation}

So far, we can solve subproblem (\ref{eq:20}) via solving subproblem (\ref{eq:22}). Due to different cases of $i$ and $j$, we summarize the solution for (\ref{eq:22}) as follows.
\begin{algorithm}[H]
	\caption{Solving subproblem (\ref{eq:22})}
	\begin{algorithmic}[S] 
		\State S0: Initialization $\hat{D}^\zeta := H + \rho \Pi_{\mathcal{K}_+^{2n}(r)} (-D^\zeta) $
		\State S1: If $i,j \in [n]$, $D^{\zeta+1}_{ij}=\| a_i - a_j \|^2$. \\
		\quad \; If $i=j,\ D^{\zeta+1}_{ii}=0,\ i \in [2n]$.\\
		\quad \; Otherwise, solve (\ref{eq:24}) with $a=-\frac{\hat{D}_{ij}^\zeta}{\rho},\ b=\frac{r}{\rho}\omega_{ij}$ to obtain $D^{\zeta+1}_{ij},\ i>j>n.$
	\end{algorithmic}
\end{algorithm}
\subsection{Solving subproblem of type (\ref{eq:24})}\label{sec:4.2}
Next, we try to derive the minimum of (\ref{eq:24}) (denoted as $\alpha^+$), due to different cases. 

\begin{figure}[htbp]
	\centering
	\subfigure[]{
		\begin{minipage}[t]{0.5\linewidth}
			\centering
			\includegraphics[width=2in]{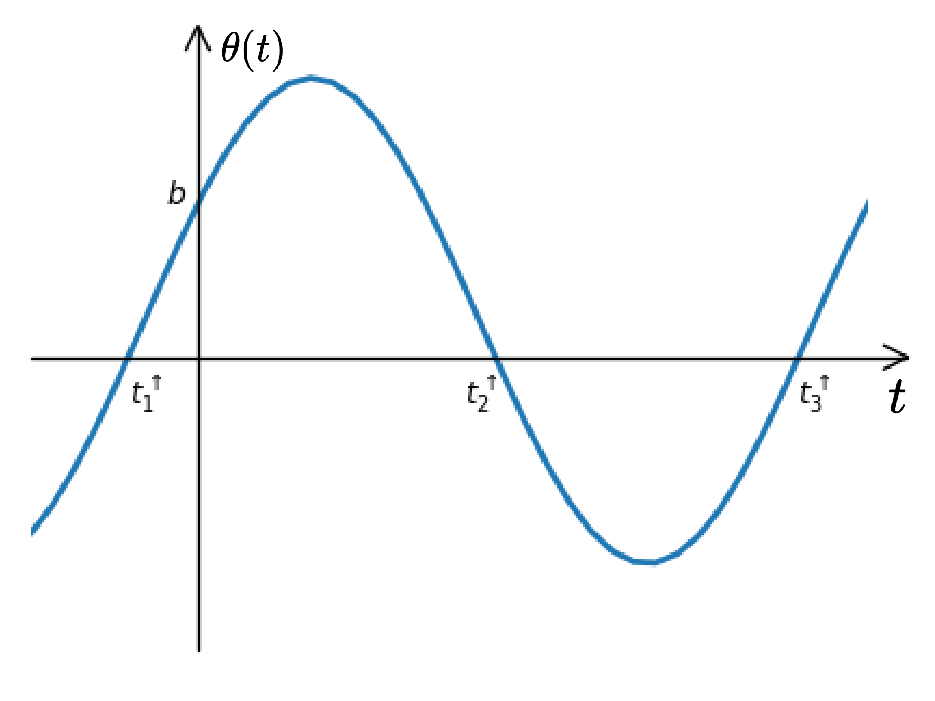}
			\label{sub1}
		\end{minipage}%
	}%
	\subfigure[]{
		\begin{minipage}[t]{0.5\linewidth}
			\centering
			\includegraphics[width=2in]{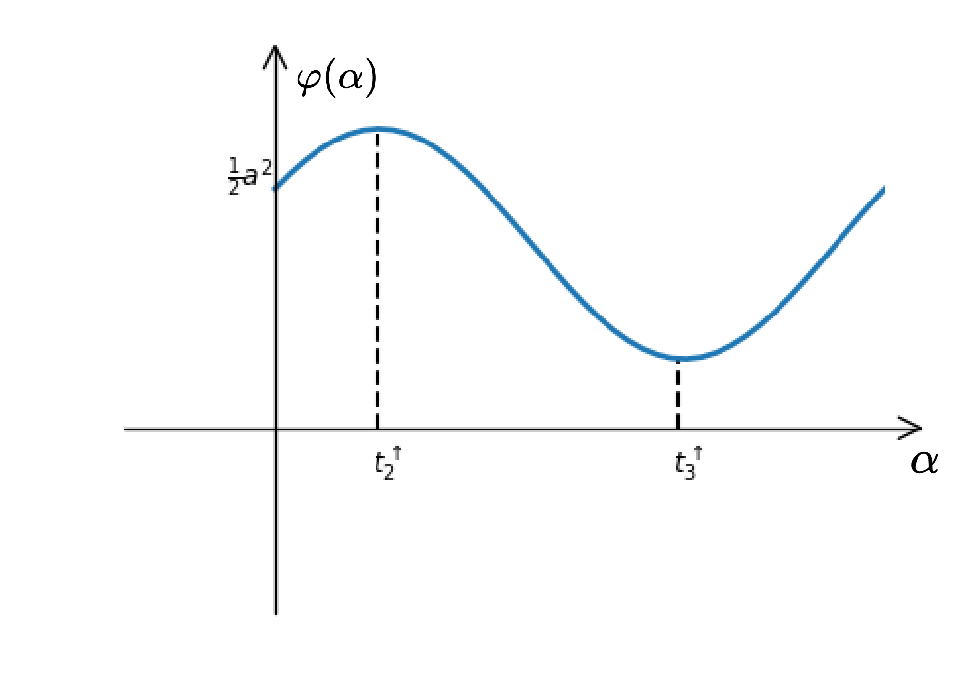}
			\label{sub2}
		\end{minipage}%
	}%
	\caption{(a) $\theta\left( t \right)$ in Case 2.1. (b) $\varphi \left( \alpha \right)$ in Case 2.2.}
\end{figure}

Recall that $b \ge 0$. For $\alpha \ge 0$ (the constraint in subproblem (\ref{eq:24})), the following holds
\begin{equation*}\label{eq:28}	
\varphi'(\alpha) = \alpha + \frac{b}{2} \frac{1}{\sqrt{\alpha}} - a.
\end{equation*}
We discuss two cases. 
\begin{enumerate}[itemindent=2em]
	\item [Case 1.] If $a \le 0$, there is $\varphi'(\alpha) > 0$ for any $\alpha > 0$. That is, $\ \varphi(\alpha)$ is an increasing function in $\alpha \in \left(0, +\infty  \right) $. Moreover, $\varphi(\alpha)$ is continuous at $\alpha = 0$. Therefore, $\ \varphi(0)$ is the minimum function value in this case. That is, $\ \alpha^+ = 0$.
	\item [Case 2.] If $a>0$, we need to investigate the sign of $\varphi'(\alpha)$ for $\alpha > 0$ in order to study the property of $\varphi(\alpha)$. To that end, rewrite $\varphi'(\alpha)$ as ($\alpha > 0$)
	$$ \varphi'(\alpha) = \frac{1}{2\sqrt{\alpha}} (2\alpha^{\frac{3}{2}} - 2a\alpha^{\frac{1}{2}} + b) := \frac{1}{2\sqrt{\alpha}}\theta(\alpha^{\frac{1}{2}}), $$
	where $\theta(t)$ is defined as $\theta(t) = 2t^3 - 2at + b$. Given the fact that $\alpha > 0$, we only need to consider the sign of $\theta(t)$ with $t>0$. Taking the gradient of $\varphi(t)$ to be zero, we can obtain 
	$$ \theta'(t) = 6t^2 - 2a = 0, $$
	giving the root of $\theta'(t)$ as $t_0=\sqrt{\frac{a}{3}} \; \left( \mbox{note that}\ a>0 \ \mbox{in this case} \right). $
	As a result, $\theta'(t) < 0$ for $t \in \left( 0,t_0 \right)$.
	Accordingly, $\theta(t)$ is monotonically decreasing at $t \in \left[ 0,t_0 \right]$ and monotonically increasing at $t \in \left[ t_0, +\infty \right]$.
	
	\item [Case 2.1] If $b \ge \frac{4}{3\sqrt{3}}a^{\frac{3}{2}}$, there is 
	$$ \theta(t_0) = 2 \left( \frac{a}{3} \right)^\frac{3}{2} - 2a \left( \frac{a}{3} \right)^\frac{1}{2} + b = -\frac{4}{3\sqrt{3}}a^{\frac{3}{2}} + b \ge 0. $$
	Therefore, for $t \in \left( 0,+\infty \right)$, $\theta(t) \ge 0$, and $\varphi(\alpha)$ is increasing over $\left( 0,+\infty \right)$. The minimum of problem (27) is then achieved at $\alpha^+=0$.
	
	\item [Case 2.2] If $a>0$ and $b < \frac{4}{3\sqrt{3}}a^{\frac{3}{2}}$, there is $\theta(t_0) < 0$. The root of $\theta(t) =0$ can be given by the Cardano formula as below. Let
	\begin{equation}\label{eq:29}
	r=\left( \frac{a}{3} \right)^{\frac{3}{2}},\; \xi = \frac{1}{3}\arccos \left(-\frac{b}{4r} \right),
	\end{equation}
	and
	\begin{equation}\label{eq:30}
	t_1 = 2\sqrt[3]{r} \cos \xi,\;
	t_2 = 2\sqrt[3]{r} \cos \left(\xi+\frac{2\pi}{3} \right), \; t_3 = 2\sqrt[3]{r} \cos \left(\xi+\frac{4\pi}{3} \right).
	\end{equation}
	Together with $\theta(0)=b \ge 0$, $\theta(t) \xrightarrow{} +\infty$ as $t \xrightarrow{} +\infty$, the typical curve of $\theta(t)$ is given as in Figure \ref{sub1}, where $t_1^{\uparrow} \le t_2^{\uparrow} \le t_3^{\uparrow}$ are basically $t_1,\ t_2,\ t_3$ after reordering in the ascending way. That is, $\theta(t) \ge 0$ for $t \in \left(0, t_2^{\uparrow} \right)$ and $t \in \left(t_3^{\uparrow}, +\infty \right)$, and $\theta(t) \le 0$ for $t \in \left(0, t_2^{\uparrow} \right) $ and $t \in \left(t_3^{\uparrow}, +\infty \right)$. As a result, $\varphi(\alpha)$ is decreasing in $t \in \left(t_2^{\uparrow}, t_3^{\uparrow} \right)$.
	Together with $\varphi(0) = \frac{1}{2}a^2 > 0 $ and $ \varphi(\alpha) \xrightarrow{} +\infty$ as $\alpha \xrightarrow{} +\infty$. The curve of $\varphi(\alpha)$ is demonstrated in Figure \ref{sub2}. Therefore, the minimum of problem (\ref{eq:24}) is achieved either at $\alpha=0$ or $\alpha=t_3^{\uparrow}$. That is, $\alpha^+ = \arg\min \limits_{\{0,t_3^{\uparrow} \}} \varphi(t)$.
	
\end{enumerate}


The above discussions lead to the following proposition.


\begin{myprop} Let $b \ge 0$. The solution of problem (\ref{eq:24}), denoted as $\alpha ^+$, is given as follows.
	\begin{enumerate}[itemindent=2em]
		\item [Case 1.] $a \le 0$, then $\alpha^+ = 0$;
		\item [Case 2.1] $a > 0$ and $b > \frac{4}{3\sqrt{3}}a^{\frac{3}{2}}$; then $\alpha^+$ = 0;
		\item [Case 2.2] $a > 0$ and $b \le \frac{4}{3\sqrt{3}}a^{\frac{3}{2}}$. By the Cardano formula, one can compute  $t_1,t_2,t_3$ by (\ref{eq:29}) and (\ref{eq:30}). 
		
		Let $\alpha_{\max} = \max\{t_1,t_2,t_3\}$. Then $ \alpha^+ = \mathop{\arg}\min\limits_{\alpha \in \{0,\ \alpha_{\max}\}}\varphi(\alpha)$. 
	\end{enumerate}
\end{myprop}

We summarize the algorithm for subproblem (\ref{eq:24}) as follows.
\begin{algorithm}[htbp]
	\caption{Algorithm for Subproblem (\ref{eq:24})}
	\begin{algorithmic}[S]
		\State S0: Initialization a:=-$\frac{\hat{D}^\zeta_{ij}}{\rho},\; b=\frac{r}{\rho}W_{ij}.$
		\State S1: If $a \le 0$ or $ a>0 $ and $b \le \frac{4}{3\sqrt{3}}a^{\frac{3}{2}} $, let $\alpha^+=0$; else, go to S2.
		\State S2: Calculate $t_1,t_2,t_3$ by (\ref{eq:29}) and (\ref{eq:30}). Define  $t_\text{max} := \max \{t_1,t_2,t_3\} $. Let $\alpha^+ = \mathop{\arg}\min\limits_{\alpha =\{0,\ t_{\max}\}} \varphi(\alpha) $
		\State S3: Output $\alpha^+$.
	\end{algorithmic}
\end{algorithm}

We end this section by the following remark.
\begin{myremar}
	From Algorithm 2 and Algorithm 3, one can see that we can obtain the explicit solution for subproblem (\ref{eq:22}). Consequently, although we deal with the matrix variable $D \in S^{2n}$ in EDM model (\ref{compactEDM}), the computational cost is not high. The subproblem in Algorithm 1 can be calculated in an elementwise way with explicit formula as in Algorithm 2 and Algorithm 3. This again verifies the advantage of our EDM model (\ref{compactEDM}) from the numerical computation point of view.
\end{myremar}


\section{Numerical Results}\label{numerical-experiments}
In this section, we test our method (denoted as MP-EDM) on some datasets from UCI Machine Learning Repository\footnote{https://archive.ics.uci.edu/ml/datasets/}. 
The experiments are conducted by using MATLAB (R2021b) on a Linux server of 256GB memory and 
52 cores with Intel(R) Xeon(R) Gold 6230R 2.1GHz CPU. Our code can be found via 
\href{}{https://www.researchgate.net/publication/376558182\_release-MPEDM}.

The parameters in MP-EDM is chosen as follows. For $w_{ij}$, a typical choice \cite{Sun} is the following $k$-nearest neighbors defined by
\begin{equation}\label{eq:3}
	\omega_{ij} = \left\{
	\begin{array}{lll}
	exp(-\varphi \| a_i - a_j \|^2) &,& \ \text { if } a_i \in k \mathrm{NN}\left(a_j\right) \text { or } a_j \in k \mathrm{NN}\left(a_i\right), \\
	0,& &\  \text{otherwise.}
	\end{array} \right.
\end{equation}
Here $k$NN$(x)$ is the set of $x$'s $k$-nearest neighbors, and $\varphi > 0$ is a constant. We choose $\varphi = 0.5$ in our experiments. 

We stop MP-EDM when the following criteria \cite{QiXiuandZhou} are satisfied
\begin{equation*}
	\text{Fprog} \le \sqrt{n} \epsilon_F \text{  and Kprog} \le \epsilon_K,
\end{equation*}
where $$\text{Fprog} = \frac{F_{\rho}(D^{\zeta-1}) - F_{\rho}(D^\zeta)}{1 + F_{\rho}(D^{\zeta-1})},\  \text{Kprog} = 1 - \frac{\sum_{i=1}^r\left[\lambda_i^2-\left(\lambda_i-\lambda^+ \right)^2 \right]}{\sum_{i=1}^n \lambda_i^2}.$$ 
Here $\lambda_1,\cdots,\lambda_r$ are the first $r$ largest eigenvalues of $D^\top$, and $\lambda_i^+ = \max \{\lambda_i, 0\}$, $ i \in [r]$. Fprog controls the convergence of $F_{\rho}(D^\zeta)$ and Kprog measures the accuracy of eigenvalues. We set $\epsilon_F=5 \times 10^{-3}$, $\epsilon_K=1 \times 10^{-3}$.

\textbf{The process of obtaining new labels.} To obtain new labels from $D^\zeta$ by solving from model EDM($r$), first we obtain $\hat{y}_1,\cdots,\hat{y}_{2n}$ by (\ref{CMDS-1}). Based on $\hat{y}_{n+1},\cdots, 
\hat{y}_{2n}$, we apply the multi-pass scheme as in ConvexClustering\footnote{https://blog.nus.edu.sg/mattohkc/softwares/convexclustering/} package of  \cite{Sun} to obtain the partition of $\{a_1,\cdots,a_n\}$, denoted as $\hat{\mathcal{V}} = \{\hat{V}_1,\cdots,\hat{V}_{\hat{K}}\}$, as presented in Algorithm \ref{alg4}, where $\hat{y}_{n+1},\cdots, \hat{y}_{2n}$ is the estimated centroid $x_1,\cdots,x_n$. We choose $\epsilon_d =	max\left( \log_2^n ,\ 10 \right) \times \epsilon_K$ in Algorithm \ref{alg4}.
\begin{algorithm}[htbp]
	\caption{Algorithm for obtaining new labels}
	\label{alg4}
	\begin{algorithmic}[1]
		\Require
		Centroids $\left[x_1,\cdots,x_n\right]$, distance tolerance $\epsilon_d$
		\State Initialize $\hat{\psi}:=\mathbf{0} \in \mathbb{R}^n,\  N:=[n],\  \hat{K}:=1$.
		
		\For {$N$ is not empty}
			\State Choose the first index $i \in N$, set $\hat{\psi}(a_i) = \hat{K}$, remove $i$ from $N$.
			\For {$j \in  N \setminus \{i\}$}
				\If {$\| x_{i} - x_{j} \| \le \epsilon_d$}
					\State $\hat{\psi}(a_j) = \hat{K}$, remove $j$ from $N$.
				\EndIf
			\EndFor
			\State $\hat{K} := \hat{K} + 1$
		\EndFor
		\Ensure Labels $\hat{\psi}$.
	\end{algorithmic}
\end{algorithm}

\textbf{Measurements for clustering}. We use Rand Index \cite{randindex} (RI) and Normalized Mutual Information \cite{Infor} (NMI) to measure whether the clustering is good or not. The bigger RI (NMI) is, the better. RI represents the ratio of pairs that are correctly clustered. Specifically, it is calculated by $\text{RI} = \frac{|S_1|+|S_2|}{\frac{n(n+1)}{2}}$, where $S_1$ is the set of pairs $a_i,a_j$ such that $\psi(a_i)=\psi(a_j)$ and $\hat{\psi}(a_i)=\hat{\psi}(a_j)$, $S_2$ is the set of pairs $a_i,a_j$ such that $\psi(a_i) \neq \psi(a_j)$ and $\hat{\psi}(a_i) \neq \hat{\psi}(a_j)$. 

Let $\hat{n}_{i} = |\hat{V}_i|$. NMI is calculated by
\begin{equation*}
\text{NMI}=\frac{\sum_{i=1}^{\hat{K}} \sum_{j=1}^{K} \left| \hat{V}_{i} \cap V_{j} \right| \log \left(\frac{n \left| \hat{V}_{i} \cap V_{j} \right|}{\hat{n}_i n_j}\right)} {\sqrt{\left(\sum_{i=1}^{\hat{K}} \hat{n}_i \log \left(\frac{\hat{n}_i}{n}\right)\right)\left(\sum_{j=1}^{K} n_j \log \left(\frac{n_j}{n}\right)\right)}}.
\end{equation*}

\subsection{Performance of parameters in MP-EDM}
\begin{figure}[hbtp]
	\centering
	\includegraphics[width=5in]{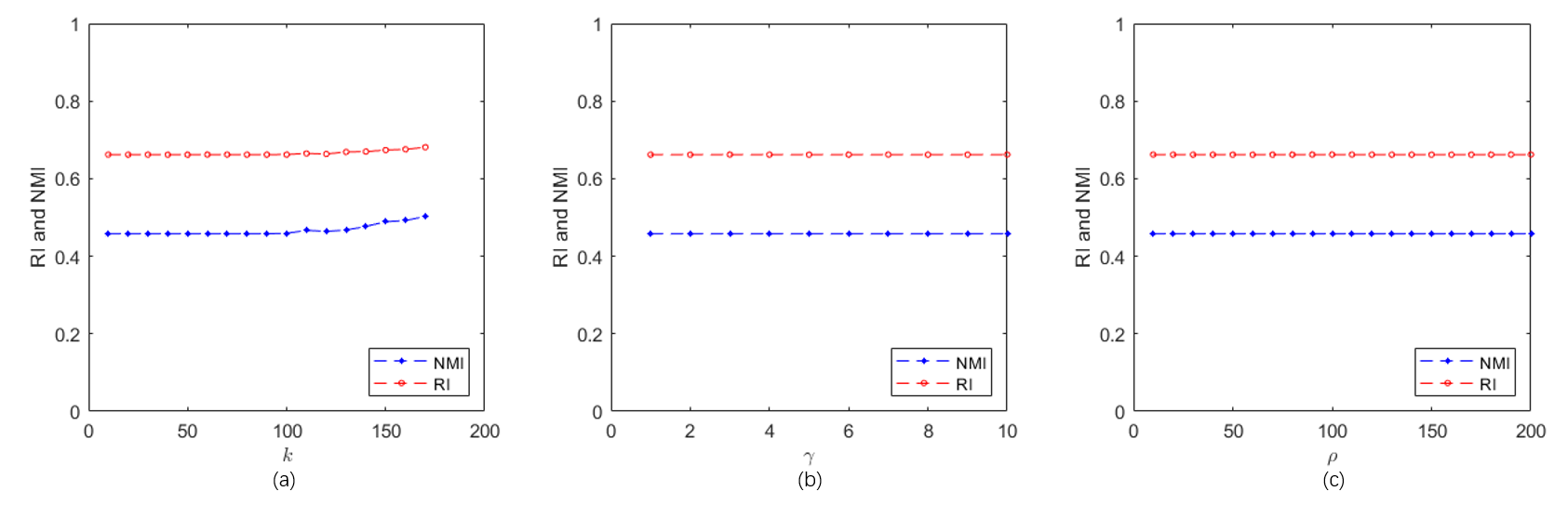}
	\caption{RI and NMI of different parameter combinations.}
	\label{testdata}
\end{figure}

In this part we test our algorithm with different combinations of parameters. Firstly we perform our algorithm on Wine, 
with $n=178,\ d=13,\ K=3$. To show how the parameters take effect on the clustering results, we set $r = 13$, $k = 50$, $\rho = 100$,  $\gamma = 2$. 
Each time we only change one single parameter among $k$, $\gamma$ and $\rho$, and compare the RI and NMI, as shown in Figure \ref{testdata}. 
As one can see from Figure \ref{testdata} (a), the choice of $k$ slightly affects the performance of clustering, and larger $k$ leads to higher RI and NMI. 
This is reasonable because large $k$ means that more neighbors are considered, which will lead to better clustering results. 
Figure \ref{testdata} (b) and (c) imply that MP-EDM is not sensitive to the choice of $\rho$ and $\gamma$. 

\begin{figure}[hbtp]
	\centering
	\includegraphics[width=5in]{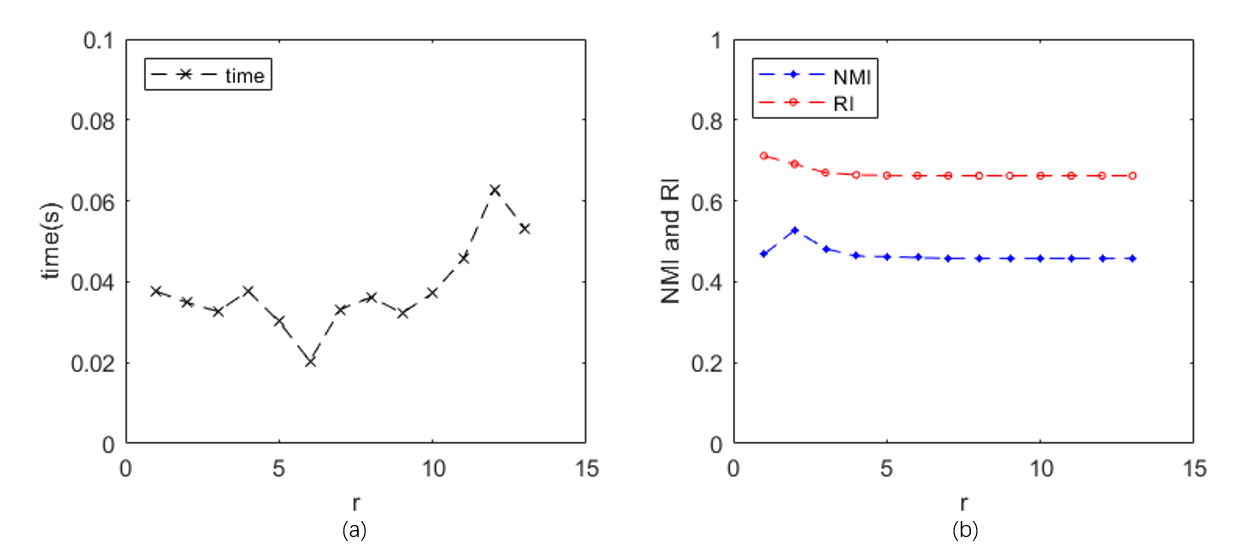}
	\caption{The cputime, RI and NMI of different $r$'s.}
	\label{wine_r}
\end{figure}

Secondly, we test the role of $r$ on Wine. We set $k = 50,\ \rho = 100,\ \gamma = 2$, and change $r=1:1:13$. 
RI, NMI and the cputime are reported in Figure \ref{wine_r}. It can be seen that for the case $r$ is much smaller than $d$, 
for example $r \le 10$, the time consumed by MP-EDM is less than that of $r \ge 11$. 
This can be explained by more computational cost calculating $\Pi_{\mathcal{K}_+^{2n}(r)}(\cdot)$ by (\ref{new1}) and (\ref{new2}) for larger $r$. 
Moreover, the resulting RI and NMI is higher when $r \le 5$, and they hardly change with $r \ge 5$. 
If we choose $r < d = 13$, we can reduce the computational cost and obtain good clustering results.

\subsection{Comparison with other methods}
Here we compare our algorithm with the popular K-Means, Spectral Clustering (SC), Hierarchical Clustering (HC)\footnote{For K-Means, SC and HC, we use built-in functions in MATLAB.} and semismooth Newton-CG augmented Lagrangian method (SSNAL) \cite{Sun2018} in ConvexClustering package on Iris, Wine, Letter-Recognition, Knowledge and MNIST. The results are reported in Table \ref{small-dataset} - Table \ref{mnistInfo}, where winners of RI and NMI are marked in bold.

From Table \ref{small-dataset} one can see that MP-EDM gives competitive RI and NMI as SSNAL on Wine and Knowledge, and MP-EDM consumes less cputime than SSNAL. MP-EDM performs worse on Iris than other algorithms, but is much faster than SSNAL.

\begin{table}[htbp]
	\centering
	\scalebox{0.9}{
	\begin{tabular}{|c|ccc|ccc|ccc|}
		\hline
		Data & \multicolumn{3}{c|}{Iris} & \multicolumn{3}{c|}{Wine} & \multicolumn{3}{c|}{Knowledge} \\ \hline
		($n,d,s,K$) & \multicolumn{3}{c|}{(150,4,4,3)} & \multicolumn{3}{c|}{(178,13,13,3)} & \multicolumn{3}{c|}{(400,5,5,4)} \\ \hline
		Methods &
		\multicolumn{1}{c|}{RI} &
		\multicolumn{1}{c|}{NMI} &
		\multicolumn{1}{c|}{Time(s)} &
		\multicolumn{1}{c|}{RI} &
		\multicolumn{1}{c|}{NMI} &
		\multicolumn{1}{c|}{Time(s)} &
		\multicolumn{1}{c|}{RI} &
		\multicolumn{1}{c|}{NMI} &
		\multicolumn{1}{c|}{Time(s)} \\ \hline
		MP-EDM &
		\multicolumn{1}{c|}{0.675} &
		\multicolumn{1}{c|}{0.479} &
		\multicolumn{1}{c|}{0.053} &
		\multicolumn{1}{c|}{0.662} &
		\multicolumn{1}{c|}{\textbf{0.457}} &
		\multicolumn{1}{c|}{0.109} &
		\multicolumn{1}{c|}{\textbf{0.728}} &
		\multicolumn{1}{c|}{\textbf{0.472}} &
		\multicolumn{1}{c|}{0.415} \\ \hline
		SSNAL &
		\multicolumn{1}{c|}{0.776} &
		\multicolumn{1}{c|}{0.761} &
		\multicolumn{1}{c|}{0.324} &
		\multicolumn{1}{c|}{0.662} &
		\multicolumn{1}{c|}{\textbf{0.457}} &
		\multicolumn{1}{c|}{0.270} &
		\multicolumn{1}{c|}{0.271} &
		\multicolumn{1}{c|}{0.000} &
		\multicolumn{1}{c|}{0.447} \\ \hline
		K-Means &
		\multicolumn{1}{c|}{0.873} &
		\multicolumn{1}{c|}{0.741} &
		\multicolumn{1}{c|}{0.003} &
		\multicolumn{1}{c|}{\textbf{0.691}} &
		\multicolumn{1}{c|}{0.424} &
		\multicolumn{1}{c|}{0.027} &
		\multicolumn{1}{c|}{0.669} &
		\multicolumn{1}{c|}{0.230} &
		\multicolumn{1}{c|}{0.004} \\ \hline
		SC &
		\multicolumn{1}{c|}{\textbf{0.892}} &
		\multicolumn{1}{c|}{\textbf{0.805}} &
		\multicolumn{1}{c|}{0.012} &
		\multicolumn{1}{c|}{0.353} &
		\multicolumn{1}{c|}{0.070} &
		\multicolumn{1}{c|}{0.018} &
		\multicolumn{1}{c|}{0.688} &
		\multicolumn{1}{c|}{0.279} &
		\multicolumn{1}{c|}{0.026} \\ \hline
		HC &
		\multicolumn{1}{c|}{0.776} &
		\multicolumn{1}{c|}{0.735} &
		\multicolumn{1}{c|}{0.001} &
		\multicolumn{1}{c|}{0.362} &
		\multicolumn{1}{c|}{0.091} &
		\multicolumn{1}{c|}{0.005} &
		\multicolumn{1}{c|}{0.278} &
		\multicolumn{1}{c|}{0.039} &
		\multicolumn{1}{c|}{0.004} \\ \hline
	\end{tabular}}
	\caption{Clustering performance on Iris, Wine and Knowledge, $r=d$}
	\label{small-dataset}
\end{table}

For Letter-Recognition, we test MP-EDM on sizes of dataset with $n$ up to 10000. One can see from Table \ref{letterInfo} that MP-EDM and SSNAL give very high RI and NMI on every $n$, but MP-EDM is faster than SSNAL. This can be explained by the fact that the two models SON($A,d$) and EDM($d$) are equivalent by Theorem \ref{theorem-1}. Although K-Means takes less time and yields a high RI, it performs poorly according to the measurement NMI.
\begin{table}[htbp]
	\centering
	\scalebox{0.7}{
	\begin{tabular}{|c|ccccc|}
		\hline
		&
		\multicolumn{5}{c|}{RI $|$ NMI $|$ Time} \\ \hline
		\multicolumn{1}{|c|}{($n,s$)} &
		\multicolumn{1}{c|}{(2000, 16)} &
		\multicolumn{1}{c|}{(4000, 16)} &
		\multicolumn{1}{c|}{(6000, 16)} &
		\multicolumn{1}{c|}{(8000, 16)} &
		\multicolumn{1}{c|}{(10000, 16)} \\ \hline
		MP-EDM &
        \multicolumn{1}{c|}{\textbf{0.961} $|$ \textbf{0.668} $|$ 04} &
		\multicolumn{1}{c|}{\textbf{0.962} $|$ \textbf{0.649} $|$ 16} &
		\multicolumn{1}{c|}{\textbf{0.962} $|$ \textbf{0.641} $|$ 34} &
		\multicolumn{1}{c|}{\textbf{0.962} $|$ \textbf{0.635} $|$ 01:05} &
		\multicolumn{1}{c|}{\textbf{0.962} $|$ \textbf{0.632} $|$ 01:01} \\ \hline
		SSNAL &
		\multicolumn{1}{c|}{\textbf{0.961} $|$ 0.657 $|$ 51} &
		\multicolumn{1}{c|}{0.961 $|$ 0.632 $|$ 02:52} &
		\multicolumn{1}{c|}{0.961 $|$ 0.620 $|$ 07:32} &
		\multicolumn{1}{c|}{0.961 $|$ 0.612 $|$ 11:51} &
		\multicolumn{1}{c|}{0.961 $|$ 0.606 $|$ 26:25} \\ \hline
		K-Means &
		\multicolumn{1}{c|}{0.927 $|$ 0.371 $|$ 0.04} &
		\multicolumn{1}{c|}{0.929 $|$ 0.371 $|$ 0.07} &
		\multicolumn{1}{c|}{0.930 $|$ 0.363 $|$ 0.09} &
		\multicolumn{1}{c|}{0.931 $|$ 0.364 $|$ 0.12} &
		\multicolumn{1}{c|}{0.931 $|$ 0.372 $|$ 0.20} \\ \hline
		SC &
		\multicolumn{1}{c|}{0.310 $|$ 0.299 $|$ 0.21} &
		\multicolumn{1}{c|}{0.149 $|$ 0.186 $|$ 0.71} &
		\multicolumn{1}{c|}{0.100 $|$ 0.135 $|$ 1.72} &
		\multicolumn{1}{c|}{0.205 $|$ 0.207 $|$ 1.64} &
		\multicolumn{1}{c|}{0.254 $|$ 0.264 $|$ 2.61} \\ \hline
		HC &
		\multicolumn{1}{c|}{0.927 $|$ 0.371 $|$ 0.04} &
		\multicolumn{1}{c|}{0.927 $|$ 0.371 $|$ 0.17} &
		\multicolumn{1}{c|}{0.927 $|$ 0.371 $|$ 0.34} &
		\multicolumn{1}{c|}{0.927 $|$ 0.371 $|$ 0.68} &
		\multicolumn{1}{c|}{0.927 $|$ 0.371 $|$ 1.10} \\ \hline
	\end{tabular}}
	\caption{Clustering performance on Letter-Recognition with $d=16$, $K=26$.}
	\label{letterInfo}%
\end{table}

In Table \ref{mnistInfo}, we test $n=2000,4000,6000,8000$, with $d=784$, $K=10$. 
We also compare the performance of MP-EDM under $r=784$ and $r < 784$. 
We use MP-EDM-1, MP-EDM-2 and MP-EDM-3 to denote MP-EDM for $r=784$, $r=s$ and $r=100$ respectively. 
MP-EDM-1, MP-EDM-2 and MP-EDM-3 perform similar to SSNAL, in terms of RI and NMI. 
Moreover, MP-EDM-3 is much faster than MP-EDM-1 and MP-EDM-2 in terms of cputime.
\begin{table}[htbp]
	\centering
	\scalebox{0.8}{
	\begin{tabular}{|c|cccc|}
		\hline
		& \multicolumn{4}{c|}{RI $|$ NMI $|$  Time}                                                                                                              \\ \hline
		\multicolumn{1}{|c|}{($n,s$)} &
		\multicolumn{1}{c|}{(2000, 601)} &
		\multicolumn{1}{c|}{(4000, 629)} &
		\multicolumn{1}{c|}{(6000, 648)} &
		\multicolumn{1}{c|}{(8000, 657)} \\ \hline
		MP-EDM-1 	  & \multicolumn{1}{c|}{0.905 $|$ 0.537 $|$ 14:08} 
					  & \multicolumn{1}{c|}{0.909 $|$ 0.559 $|$ 09:33}   
					  & \multicolumn{1}{c|}{0.897 $|$ \textbf{0.523} $|$ 03:43}    
					  & \multicolumn{1}{c|}{0.898 $|$ \textbf{0.511} $|$ 05:57}    \\ \hline
		MP-EDM-2 	  & \multicolumn{1}{c|}{0.905 $|$ 0.538 $|$ 30:24}   
					  & \multicolumn{1}{c|}{\textbf{0.910} $|$ \textbf{0.561} $|$ 08:32}       
					  & \multicolumn{1}{c|}{0.896 $|$ 0.519 $|$ 05:02}    
					  & \multicolumn{1}{c|}{0.898 $|$ 0.508 $|$ 09:42}    \\ \hline
		MP-EDM-3 	  & \multicolumn{1}{c|}{\textbf{0.908} $|$ \textbf{0.574} $|$ 16}   
			 		  & \multicolumn{1}{c|}{0.898 $|$ 0.519 $|$ 48}       
			 		  & \multicolumn{1}{c|}{0.874 $|$ 0.465 $|$ 02:06}    
			 		  & \multicolumn{1}{c|}{0.868 $|$ 0.439 $|$ 03:49}    \\ \hline
		SSNAL         & \multicolumn{1}{c|}{0.899 $|$ 0.549 $|$ 33:35} 
					  & \multicolumn{1}{c|}{0.899 $|$ 0.526 $|$ 01:51:25} 
					  & \multicolumn{1}{c|}{\textbf{0.899} $|$ 0.514 $|$ 04:07:16} 
					  & \multicolumn{1}{c|}{\textbf{0.900} $|$ 0.506 $|$ 06:21:16} \\ \hline
		K-Means       & \multicolumn{1}{c|}{0.885 $|$ 0.502 $|$ 0.23}    
					  & \multicolumn{1}{c|}{0.888 $|$ 0.502 $|$ 0.31}       
					  & \multicolumn{1}{c|}{0.892 $|$ 0.516 $|$ 0.49}       
					  & \multicolumn{1}{c|}{0.878 $|$ 0.489 $|$ 0.52}       \\ \hline
		SC            & \multicolumn{1}{c|}{0.310 $|$ 0.299 $|$ 0.37}    
					  & \multicolumn{1}{c|}{0.149 $|$ 0.186 $|$ 0.75}       
					  & \multicolumn{1}{c|}{0.100 $|$ 0.135 $|$ 1.59}       
					  & \multicolumn{1}{c|}{0.205 $|$ 0.207 $|$ 1.38}       \\ \hline
		HC            & \multicolumn{1}{c|}{0.074 $|$ 0.090 $|$ 0.14}    
					  & \multicolumn{1}{c|}{0.059 $|$ 0.067 $|$ 0.32}       
					  & \multicolumn{1}{c|}{0.051 $|$ 0.051 $|$ 0.73}       
					  & \multicolumn{1}{c|}{0.061 $|$ 0.078 $|$ 0.81}       \\ \hline
	\end{tabular}}
	\caption{Clustering performance on MNIST with $d=784$ and $K=10$.}
	\label{mnistInfo}
\end{table}

\section{Conclusions}\label{sec:6}
In this paper, we proposed a Euclidean distance matrix model based on the SON model for clustering. An efficient majorization penalty algorithm was proposed to solve the resulting model. The exact recovery property of the EDM model is established under some assumptions. Extensive numerical experiments were conducted to demonstrate the efficiency of the proposed model and the majorization penalty algorithm. Notice that in Section 2, if $r < s$, $a_1,\cdots,a_n$ can not be embedded into $r$-dimensional space. In this case, it is not clear whether the exact recovery property of EDM($r$) maintains or not. We will continue to investigate this question in future.

\section*{Acknowledgments}
We would like to thank Professor Zaid Harchaoui for handling our submission and the anonymous reviewers for 
the wonderful comments, based on which we improved our paper. We would also like to thank Dr. Yancheng Yuan 
from Hong Kong Polytechnic University for the insightful discussions on the exact recovery of our EDM model.

\section*{Appendix}

Proof of Theorem \ref{Theo5Appendix} is presented in this part.
\begin{proof}
	Recall $B$ defined in (\ref{eq:0}) $$B:= \left\{D \in S^{2n} \;|\; diag(D)=0,\ D_{ij} = \| a_i-a_j \|^2,\ i,j\in [n] \right\}.$$
	Since $D_\rho^*$ is an optimal solution of (\ref{model with rhog}), we have $ D_\rho^* \in B$. Given $g\left(D\right) = 0,\ f\left(D_\rho^*\right) \geq 0$,
	\begin{equation*}
	f\left(D\right)= f\left(D\right)+\rho g\left(D\right) \geq f\left(D_\rho^*\right)+\rho g\left(D_\rho^*\right) \geq \rho g\left(D_\rho^*\right).
	\end{equation*}
	Therefore, we have
	\begin{equation*}
	g\left(D_\rho^*\right) \leq \frac{f\left(D\right)}{\rho} \le \frac{f\left(D\right)}{\rho_\epsilon} = \epsilon.
	\end{equation*}
	Since $g\left(D^*\right) = 0,\ g\left(D_\rho^*\right) \geq 0$, we have
	\begin{equation*}
	f\left(D^*\right)= f\left(D^*\right)+\rho g\left(D^*\right) \geq f\left(D_\rho^*\right)+\rho g\left(D_\rho^*\right)	\geq f\left(D_\rho^*\right). \quad \qed
	\end{equation*}
\end{proof}

%
%

\bibliographystyle{spmpsci}       
\bibliography{sample}   

\begin{thebibliography}{10}
\providecommand{\url}[1]{{#1}}
\providecommand{\urlprefix}{URL }
\expandafter\ifx\csname urlstyle\endcsname\relax
  \providecommand{\doi}[1]{DOI~\discretionary{}{}{}#1}\else
  \providecommand{\doi}{DOI~\discretionary{}{}{}\begingroup
  \urlstyle{rm}\Url}\fi

\bibitem{BaiandQi}
Bai, S.H., Qi, H.D.: Tackling the flip ambiguity in wireless sensor network
  localization and beyond.
\newblock Digital Signal Processing \textbf{55}(C), 85--97 (2016).
\newblock \doi{10.1016/j.dsp.2016.05.006}

\bibitem{Biswas}
Biswas, P., Liang, T.C., Toh, K.C., Ye, Y., Wang, T.C.: Semidefinite
  programming approaches for sensor network localization with noisy distance
  measurements.
\newblock IEEE Transactions on Automation Science and Engineering
  \textbf{3}(4), 360--371 (2006).
\newblock \doi{10.1109/TASE.2006.877401}

\bibitem{Mordern}
Borg, I., Groene, P.J.F.: Modern Multidimensional Scaling.
\newblock Springer, Berlin (2005)

\bibitem{ChiandLange}
Chi, E.C., Lange, K.: Splitting methods for convex clustering.
\newblock Journal of Computational and Graphical Statistics \textbf{24}(4),
  994--1013 (2015).
\newblock \doi{10.1080/10618600.2014.948181}

\bibitem{Cox}
Cox, M.A.A., Cox, T.F.: Multidimensional Scaling.
\newblock Springer Berlin Heidelberg, Berlin, Heidelberg (2008).
\newblock \doi{10.1007/978-3-540-33037-0\_14}

\bibitem{Dattorro}
Dattorro, J.: Convex Optimization and \hbox{E}uclidean Distance Geometry.
\newblock Meboo Publishing, USA (2005)

\bibitem{DingandQi}
Ding, C., Qi, H.D.: Convex optimization learning of faithful \hbox{E}uclidean
  distance representations in nonlinear dimensionality reduction.
\newblock Mathematical Programming \textbf{164}(1), 341--381 (2017).
\newblock \doi{10.1007/s10107-016-1090-7}

\bibitem{Fiedler1973}
Fiedler, M.: Algebraic connectivity of graphs.
\newblock Czechoslovak Mathematical Journal \textbf{23}(2), 298--305 (1973)

\bibitem{YandGao}
Gao, Y.: Structured low rank matrix optimization problems: \hbox{A} penalty
  approach.
\newblock Ph.D. thesis, National University of Singapore (2010)

\bibitem{gower2004}
Golub, G.H., Loan, C.V.: Matrix computations.
\newblock Johns Hopkins University Press, Baltimore (1996)

\bibitem{Hocking}
Hocking, T.D., Joulin, A., Bach, F., Vert, J.P.: Clusterpath an algorithm for
  clustering using convex fusion penalties.
\newblock 28th International Conference on Machine Learning pp. 341--381 (2011)

\bibitem{randindex}
Hubert, L., Arabie, P.: Comparing partitions.
\newblock Journal of Classification \textbf{2}(1), 193--218 (1985).
\newblock \doi{10.1007/BF01908075}

\bibitem{LiandQi}
Li, Q.N., Qi, H.D.: An inexact smoothing \hbox{N}ewton method for
  \hbox{E}uclidean distance matrix optimization under ordinal constraints.
\newblock Journal of Computational Mathematics \textbf{35}(4), 469--485 (2017).
\newblock \doi{10.4208/jcm.1702-m2016-0748}

\bibitem{2011Block}
Li, Q.N., Qi, H.D., Xiu, N.H.: Block relaxation and majorization methods for
  the nearest correlation matrix with factor structure.
\newblock Computational Optimization and Applications \textbf{50}(2), 327--349
  (2011).
\newblock \doi{10.1007/s10589-010-9374-y}

\bibitem{Lindsten}
Lindsten, F., Ohlsson, H., Ljung, L.: Clustering using sum-of-norms
  regularization: With application to particle filter output computation.
\newblock In: IEEE Statistical Signal Processing Workshop, pp. 201--204 (2011).
\newblock \doi{10.1109/SSP.2011.5967659}

\bibitem{LuandLi}
Lu, S.T., Zhang, M., Li, Q.N.: Feasibility and a fast algorithm for
  \hbox{E}uclidean distance matrix optimization with ordinal constraints.
\newblock Computational Optimization and Applications \textbf{76}(2), 535--569
  (2020).
\newblock \doi{10.1007/s10589-020-00189-9}

\bibitem{Pelckmans}
Pelckmans, K., Brabanter, J.D., Suykens, J., Moor, B.D.: Convex clustering
  shrinkage.
\newblock In: PASCAL Workshop on Statistics and Optimization of Clustering
  Workshop (2005)

\bibitem{Qi2013}
Qi, H.D.: A semismooth \hbox{N}ewton's method for the nearest \hbox{E}uclidean
  distance matrix problem.
\newblock SIAM Journal on Matrix Analysis and Applications \textbf{34}(34),
  67--93 (2013).
\newblock \doi{10.1137/110849523}

\bibitem{Qi2013A}
Qi, H.D., Xiu, N.H., N, Yuan, X: A \hbox{L}agrangian dual approach to the
  single source localization problem.
\newblock IEEE Transactions on Signal Processing \textbf{61}(15), 3815--3826
  (2013).
\newblock \doi{10.1109/TSP.2013.2264814}

\bibitem{QiandYuan2014}
Qi, H.D., Yuan, X.M.: Computing the nearest \hbox{E}uclidean distance matrix
  with low embedding dimensions.
\newblock Mathematical Programming \textbf{147}(1-2), 351--389 (2014).
\newblock \doi{10.1007/s10107-013-0726-0}

\bibitem{schoen}
Schoenberg, I.J.: Metric spaces and positive definite functions.
\newblock Transactions of the American Mathematical Society \textbf{44}(3),
  522--536 (1938)

\bibitem{Sun}
Sun, D.F., Toh, K.C., Yuan, Y.: Convex clustering: Model, theoretical guarantee
  and efficient algorithm.
\newblock Journal of Machine Learning Research \textbf{22}(9), 1--32 (2021).
\newblock \doi{10.5555/3546258.3546267}

\bibitem{Toh2008}
Toh, K.C.: An inexact primal-dual path-following algorithm for convex quadratic
  \hbox{SDP}.
\newblock Mathematical Programming \textbf{112}(1), 221--254 (2008).
\newblock \doi{10.1007/s10107-006-0088-y}

\bibitem{Infor}
Vinh, N.X., Epps, J., Bailey, J.: Information theoretic measures for
  clusterings comparison: Variants, properties, normalization and correction
  for chance.
\newblock Journal of Machine Learning Research \textbf{11}(95), 2837--2854
  (2010).
\newblock \doi{10.5555/1756006.1953024}

\bibitem{von2007tutorial}
Von~Luxburg, U.: A tutorial on spectral clustering.
\newblock Statistics and computing \textbf{17}, 395--416 (2007).
\newblock \doi{10.1007/s11222-007-9033-z}

\bibitem{yuan2023randomly}
Wang, Z.W., Yuan, Y.C., Ma, J.M., Zeng, T.Y., Sun, D.F.: Randomly projected
  convex clustering model: Motivation, realization, and cluster recovery
  guarantees (2023).
\newblock \urlprefix\url{https://arxiv.org/abs/2303.16841}

\bibitem{Yao2020}
Yao, Z.Q., Dai, Y.J., Li, Q.N., Xie, D., Liu, Z.H.: A novel posture positioning
  method for multi-joint manipulators.
\newblock IEEE Sensors Journal \textbf{20}(23), 14310--14316 (2020).
\newblock \doi{10.1109/JSEN.2020.3007701}

\bibitem{Sun2018}
Yuan, Y., Sun, D.F., Toh, K.C.: An efficient semismooth {N}ewton based
  algorithm for convex clustering.
\newblock In: 35th International Conference on Machine Learning, vol.~13, pp.
  9085--9095 (2018)

\bibitem{ZhaiandLi}
Zhai, F.Z., Li, Q.N.: A \hbox{E}uclidean distance matrix model for protein
  molecular conformation.
\newblock Journal of Global Optimization \textbf{76}(4), 709--728 (2020).
\newblock \doi{10.1007/s10898-019-00771-4}

\bibitem{QiXiuandZhou}
Zhou, S.L., Xiu, N.H., Qi, H.D.: A fast matrix majorization-projection method
  for constrained stress minimization in \hbox{MDS}.
\newblock IEEE Transactions on Signal Processing \textbf{66}(3), 4331--4346
  (2018).
\newblock \doi{10.1109/TSP.2018.2849734}

\bibitem{ZhouandQi}
Zhou, S.L., Xiu, N.H., Qi, H.D.: Robust \hbox{E}uclidean embedding via
  \hbox{EDM} optimization.
\newblock Mathematical Programming Computation \textbf{12}(3), 337--387 (2019).
\newblock \doi{10.1109/TSP.2018.2849734}

\end{thebibliography}


\end{document}